      \documentclass{elsarticle}
      \usepackage{hyperref}
      
            \hypersetup{colorlinks,
	      citecolor=blue, 
	      filecolor=black,
	      linkcolor=black,
	      urlcolor=blue   
          }
          
      \journal{Journal of Food Engineering}
      \usepackage{graphics}
      \usepackage[utf8x]{inputenc}		
      \usepackage{amsmath}                     
      \usepackage{amssymb}                     
      \usepackage{multirow}			

      \DeclareFontFamily{U}{euc}{}
      \DeclareFontShape{U}{euc}{m}{n}{<-6>eurm5<6-8>eurm7<8->eurm10}{}%
      \DeclareSymbolFont{AMSc}{U}{euc}{m}{n} 
      \DeclareMathSymbol{\umu}{\mathord}{AMSc}{"16} 

      \newcommand{\umum}{\hspace{0.6 mm}\,\umu\textrm{m}}

\begin{document}
      \begin{frontmatter}
      

\title{
      Quantitative in-situ monitoring of germinating barley seeds using X-ray dark-field radiography
      }

\author[mymainaddress]{Mikkel Schou Nielsen \corref{mycorrespondingauthor}}
\cortext[mycorrespondingauthor]{Corresponding author}
\ead{schou@nbi.ku.dk}

\author[mysecondaryaddress]{Kasper Borg Damkj\ae r}

\author[mymainaddress]{Robert Feidenhans'l}

\address[mymainaddress]{Niels Bohr Institute, University of Copenhagen, Universitetsparken 5, 2100 Copenhagen, Denmark}
\address[mysecondaryaddress]{FOOD, University of Copenhagen, Rolighedsvej 30, 2000 Frederiksberg, Denmark}

    \begin{abstract}
    During production of malt from barley seeds, cell walls and starch granules in the endosperm are degraded. Although this modification process is important for malt quality, the modification patterns of individual barley seeds have yet to be reported. The use of destructive microscopy methods have previously limited the investigations to ensemble averages.

    X-ray dark-field radiography is a recent non-destructive imaging method which is sensitive to microstructural variations. In this study, the method was applied for quantitative in-situ monitoring of barley seeds. Microstructural changes relating to water uptake and modification were monitored over a 43–55 h period.

    Sub-resolution stress cracks as well as a dark-field signal believed to originate from starch granules were detected. The evolution of the dark-field signal followed the known modification pattern in barley seeds. Through image analysis, quantitative parameters describing the movement of the front of the observed pattern were obtained.

    Based on these findings, X-ray dark-field radiography presents a possible novel approach to monitor the modification of germinating barley seeds.
    \end{abstract}



      
      \end{frontmatter}

\section{Introduction}
 High quality malt from barley seeds is crucial in the production of high quality beer and malt whisky. In the production of malt, uptake of water during steeping is followed by enzymatic degradation of the barley endosperm during germination. The degradation process, known as modification, causes structural changes in the barley endosperm as cell walls, starch granules and their surrounding protein matrix are partially hydrolysed by enzymes. The degradation initiates next to the scutellum and advances towards the scutellar face followed by endosperm degradation beneath the aleurone layer. Usually, this advancement happens parallel to the scutellar face although other patterns can be observed \citep{briggs_patterns_1983,briggs_malt_2002,gianinetti_theoretical_2009}. 

The time needed for modification depends on the speed of the degradation front \citep{gianinetti_theoretical_2009} and can vary between different barley cultivars \citep{brennan_cultivar_1997}. Theoretical enzyme kinetic models predict that the front will move parallel to the scutellar face at a constant speed \citep{obrien_modification_2005, fowkes_application_2010}. However, experimental investigations are needed to verify these results \citep{fowkes_application_2010}.

Although the modification over time has been investigated extensively using microscopy methods \citep{aastrup_rapid_1981, briggs_patterns_1983, brennan_cultivar_1997, ferrari_constitutive_2010} so far only ensemble averages have been obtained. Since the barley seeds are sectioned and stained using e.g. calcofluor \citep{aastrup_quantitative_1980, aastrup_rapid_1981} before photographed, the applied methods are destructive in nature. Conversely, in order to follow the modification process in individual seeds, a non-destructive imaging method would be needed. 

X-ray dark-field radiography is a novel, non-destructive imaging technique introduced by \citep{pfeiffer_hard_dark_field}. It uses an X-ray grating interferometer \citep{david,momose_grating,weitkamp_grating} which can be adapted to laboratory-based setups \citep{pfeiffer_tube}.Whereas contrast in conventional transmission radiography is based on X-ray attenuation, the dark-field signal originates from ultra small-angle X-ray scattering of microstructures within the material. Thus, dark-field radiography is sensitive to structural differences on the micrometer scale \citep{bech_quantitative_2010, yashiro_origin_2010, lynch_interpretation_2011} which makes it a complementary image modality to conventional radiography. In addition, sub-resolution edges may be detected \citep{yashiro_unresolvable_2015, lauridsen_fractures_2015}. Within food science, the structural sensitivity of dark-field radiography has been used for foreign body detection of paper and insects in food products \citep{nielsen_fb_2013}. Furthermore, microstructural differences between fresh, frozen and defrosted fruit and berries were recently distinguished \citep{nielsen_frozen_2014}. 

Several structures on the micrometer scale are present in the barley seeds. The thickness of the barley cell walls in the aleurone layer and endosperm ranges from $0.5-4 \umum$ , \citep{fincher_1975, gram_ii_1982, lazaridou_morphology_2008, palmer_1998}. Endosperm starch granules typically range from $0.5-48 \umum$ in diameter \citep{ao_granules_2007, briggs_barley_book, lindeboom2004, macgregor1979, stoddard_starch_1999} depending on factors such as barley variety and growth conditions \citep{briggs_barley_book}. In accordance, since the X-ray dark-field signal is sensitive to microstructural changes, degradation of barley microstructures resulting from endosperm modification could be detectable using X-ray dark-field radiography.

The aim of this study was to apply X-ray dark-field radiography to quantitatively investigate and monitor microstructural changes relating to water uptake and modification  in barley seeds during a 48 hour germination period. The X-ray dark-field radiographs were compared to conventional X-ray transmission radiographs.

\section{Materials and methods}

\subsection{X-ray grating interferometry}\label{sec_gbi}

\begin{figure}[tb]
  \centering
  \includegraphics[width=0.9\textwidth]{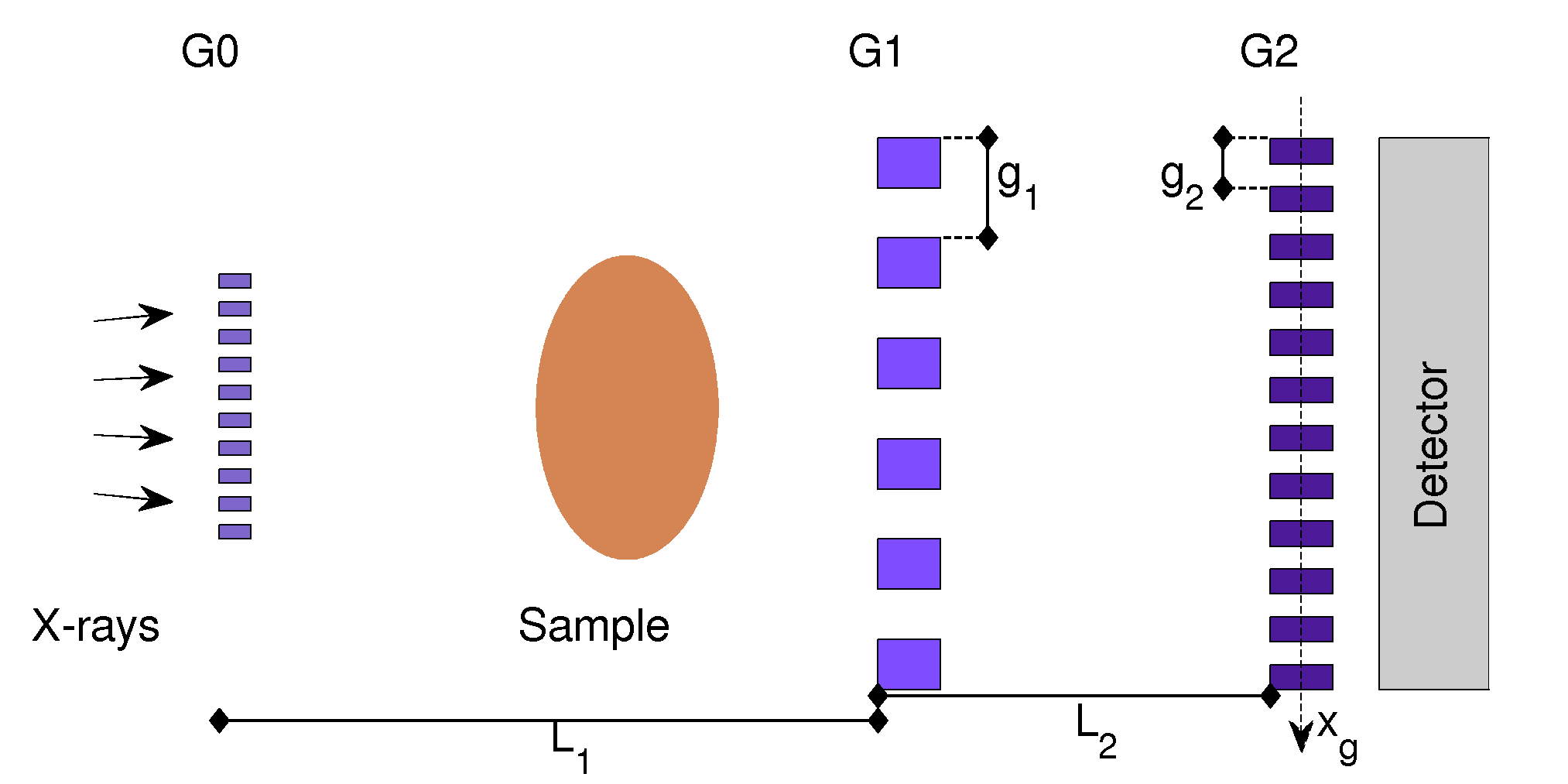}
  \caption{ \label{fig_setup}
  \footnotesize{A sketch of the X-ray grating interferometer setup with distances indicated. The phase-grating G1 creates an intensity pattern which is analyzed by the analyzer grating G2. The spatial coherence is assured by the source grating G0.}
  }
  \end{figure}

The X-ray grating interferometer has previously been described in detail \citep{pfeiffer_tube, weitkamp_grating}. As seen in figure \ref{fig_setup}, it consists of an X-ray phase-grating G1 and an analyzer absorption-grating G2. At laboratory setups a third grating, G0, is included to obtain satisfactory spatial coherence in the horizontal direction perpendicular to the grating lines.

An interference pattern is generated by G1 and creates a periodic intensity modulation at the position of G2. The latter is used to analyse the position, mean value and amplitude of the intensity modulation by moving one of the gratings in steps through the period of the pattern while recording an exposure at each step. In addition to a sample phase-stepping scan, a reference scan without the sample is carried out. The presence of a sample distorts the interference pattern and the distortion can be calculated by analysing the intensity modulations with and without the sample present \citep{pfeiffer_x-ray_2009}. From these calculations, the transmission and dark-field radiographs can be extracted with respect to the background. 

Following \citep{strobl_general_2014}, the microstructure sizes that the dark-field contrast of a grating interferometer is sensitive to may be gauged by introducing the autocorrelation length, $d_{\textrm{GI}}$, of the setup:

\begin{align}
d_{\textrm{GI}}=\frac{\lambda}{g_2\cdot L_s} 
\end{align}

where $\lambda$ is the X-ray wavelength, $g_2$ is the grating period of G2, and $L_s$ relates to the position of the sample in the interferometer. When the sample is placed between G0 and G1, $L_s$ is given by $L_s=(L_1-z_s)/L_1\cdot L_2$ where $L_1$ ($L_2$) is the distance between G0 and G1 (G1 and G2) and $z_s$ is the distance of the sample upstreams to G1. 

For spherical microparticles in a matrix, an analytical expression for the dark-field signal has been derived \citep{yashiro_origin_2010,lynch_interpretation_2011,strobl_general_2014}. In this case, the microstructures that contribute to the dark-field signal are in the size range of $0.5\cdot d_\textrm{GI}$ to $15\cdot d_{\textrm{GI}}$.

So far, the wavelength-dependent $d_{\textrm{GI}}$ has only been defined for a monochromatic X-ray beam. As an approximation in the case of a white X-ray beam, an effective energy could be introduced to calculate an effective $d_{\textrm{GI}}$.

\subsection{Barley samples and preparation}

Barley samples of malting-quality (Hordeum vulgare L. cv. Odyssey) grown 2014 were obtained from the Danish Malting Group A/S (DMG). The barley had a moisture content of 13.5\%, a protein content (dry matter) of 9.7\% and a grading of 1.9\%  $<$  2.2 mm and 95.9\% $>$ 2.5 mm (analyses done by DMG). The barley was stored in an airtight container at 5$^\circ$C. 

The germination energy of $4\, \times \, 100$ seeds was determined over 72 h using the BRF method (EBC method 3.6.2) \citep{analytica_ebc}. Two 90 mm filter papers were placed in the bottom of each 90 mm Petri dish, and 4 ml of water was added.

For the X-ray dark-field measurements, a modified version of the BRF method (EBC method 3.6.2) was used to standardize the barley germination.  Five seeds with the ventral side facing the wetted filter papers were placed on a line in the centre of each Petri dish. Four Petri dishes prepared as stated above were used in total, and the seeds were numbered from 1 to 20. The seeds were not removed during the experiment which lasted 43-55 hours. The total measurement time was limited by the X-ray setup.

\subsection{Experimental X-ray grating interferometer setup}\label{sec_nbi_interferometer}
     
The experiments were done at the Niels Bohr Institute using an experimental X-ray rotating anode tube setup with a grating interferometer. The Rigaku rotation anode tube had a copper target and was set at an acceleration voltage of 40 kV and a filament current of 50-70 mA. The effective source size was 1 mm $\times$ 1 mm.
The interferometer used a $\pi$ phase-grating for G1 with a period of $3.5 \umum$, and G0 and G2 gratings with periods of $14.1 \umum$ and $2.0 \umum$, respectively. The grating design energy was 28 keV.  The interferometer was set up with a G0-to-G1 distance of 139 cm and a G1-to-G2 distance of 20 cm. The sample stage was positioned 15 cm upstreams of G1. Using the listed distances and a design energy of 28 keV, this gives an effective $d_{\textrm{GI}}= 4 \umum$.  The images were recorded with a PILATUS 100k detector with 195x487 pixels and an effective pixel size at the sample of $134 \umum \times  134 \umum$.

\subsection{X-ray radiography measurements}

Two measurement time series A and B were performed. In both, a stack of two Petri dishes each containing 5 seeds placed on a line were imaged using X-ray grating-based radiography. Seeds 1 to 10 were measured in time series A and 11 to 20 in B. The total duration was 43 h and 55 h, respectively, which was limited by the long-time stability of the X-ray rotating anode.

At intervals of ten minutes during the time series, a phase-stepping scan of the sample was conducted. In addition, prior to each time-series five reference phase-stepping scans were conducted as described in section \ref{sec_gbi}. For sample and reference scans, 16 phase-steps were used in both series A and B. 

In an attempt to increase the long-time stability of the setup, the filament current was reduced from 70 mA in series A to 50 mA in series B. To ensure comparable detector counts and X-ray dose, the exposure time was adjusted accordingly. Thus, the 10 s exposure per acquisition used in A was adjusted to 14 s in B which gave a time per phase-stepping scan of 3 and 4 min, respectively. The measurements were carried out with air as background using the setup described in section \ref{sec_nbi_interferometer}. The temperature and humidity for the two measurements were similar with a temperature of 19 degrees C. From the acquired phase-stepping scans, X-ray transmission and dark-field radiographs were calculated.

After the measurement, it was observed that all seeds except for seed number 2 and 19 had visible rootlets which is a sign of germination. Thus, seed number 2 and 19 might not have germinated.

\subsection{Dose calculation}

The dose d imparted on the barley seeds was estimated using the expression $d=N\cdot E \cdot \mu/\rho$ where $\mu/\rho$ is the mass attenuation coefficient, $N$ is the number of incident photons per unit area and E is the photon energy \citep{Howells2009}. In place of the polychromatic energy spectrum, E was approximated as an effective energy of 25 keV. Assuming the attenuation of barley seeds to be roughly the same as for water, this gave $\mu/\rho = 0.53\, \textrm{cm}^2/\textrm{g}$.

The number of photons $N =1.3\cdot10^10 \textrm{photons}/\textrm{cm}^2$ was estimated as follows. First, the total number of photons detected in an empty area over the 16 phase-steps was estimated. From this, the number of incident photons on the $320 \umum$ thick Si detector chip was calculated by dividing with the detection efficiency of 16\% at 25 keV. Finally, $N$ was found by multiplying by a factor of two since 50\% of the number of photons incident on the sample were assumed to be absorbed by the gratings. 

Finally, this gave a dose per radiograph of $d= 28 \, \textrm{mGy}$ and a total dose for the two time series of roughly 6 Gy and 9 Gy respectively.

\subsection{Image analysis}

In order to monitor the barley seeds quantitatively, image analysis was applied to the radiographs. This aimed at locating the position and the area of the measured scattering signal of the dark-field radiographs.

Initially, a noise reduction was applied to the transmission and dark-field radiographs, and a mask for each barley seed was created. For the noise reduction, a median filter with a 2x2x2 support kernel was applied to the full time series. To create the mask, a bivariate threshold segmentation followed by a morphological opening were applied to the transmission and dark-field radiographs. A lower transmission value of 0.4 and an upper value of 0.8 were used. A dark-field threshold of 0.8 was used.

To locate the front of the scattering signal, edge detection was applied to the masked dark-field radiographs. The internal border of the scattering signal was detected using derivatives of a Gaussian filter. By convoluting with first, second and third order derivatives of a Gaussian filter along the major axis, the edges were found by looking at zero crossings in the second order derivative. A $21 \times 21$ support kernel and a variance of 4 pixels were used for the Gaussian filter. The position of the front was recorded in intervals of 30 min using every third radiograph. The position was measured with respect to a line through the center-of-mass of the seed and along the major axis.

The area of the barley seeds in the radiographs was calculated from the mask. An additional threshold segmentation of the masked seeds in the dark-field radiograph was done to extract the area with a scattering signal. A dark-field threshold value of 0.4 was used. Finally, the ratio of the two areas was calculated.

\section{Results}

\subsection{Germination energy}

Table \ref{tab_germinate} display the data for a four-time determination of the germination energy.  The average germination energy was 98\%. On average  20\% and 83\% of the barley seeds were chitted within 24 hours and 48 hours. Normally, the percentage after 24 hours is higher. However, it varies between barley types.

\begin{table}[h!tb]
\centering
\renewcommand{\tabcolsep}{2.5mm}
\renewcommand{\arraystretch}{1.2}
\begin{tabular}{l|c|c|c|c}
\multicolumn{5}{c}{\textbf{Germination energy }}\\
\multicolumn{5}{c}{(cumulated) [\%]}\\
\hline
 & A & B & C & D \\
\hline
24 hours & 15 & 20 & 21& 24\\
48 hours & 78 & 87 & 83 & 84\\
72 hours & 97 & 99 & 98 & 97\\
\hline
Cumulated [\%]& 97 &99 & 98& 97\\
\hline
\end{tabular}
\caption{\label{tab_germinate} Four-time determination of germination energy (Analytica-EBC 3.6.3) for barley variety Odyssey harvested in 2014.}
\end{table}

\subsection{X-ray radiography of germinating barley seeds}

\begin{figure}[tb]
  \centering
  \includegraphics[width=0.9\textwidth]{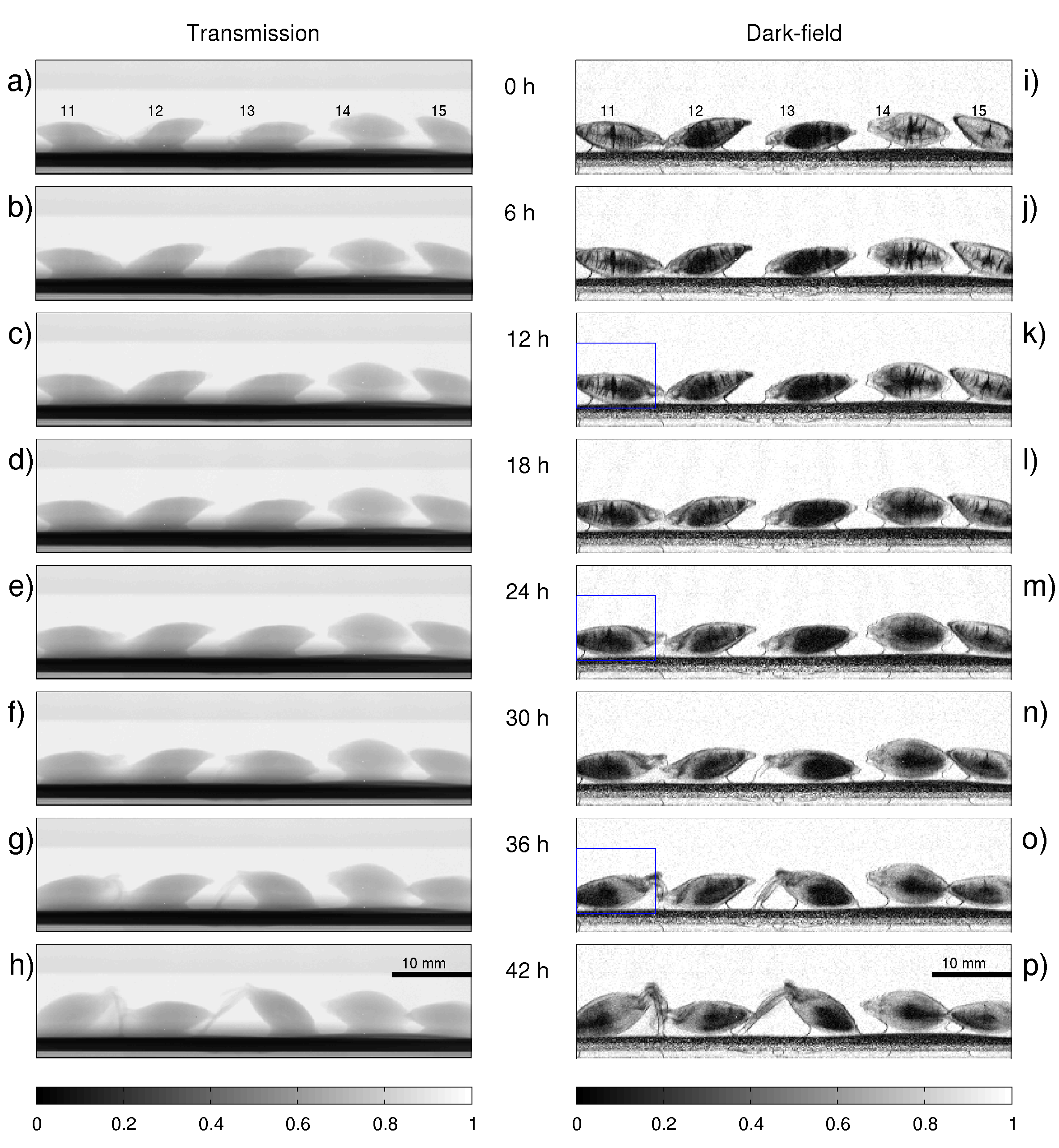}
  \caption{ \label{fig_trans_df}
  \footnotesize{Transmission and dark-field radiographs of barley seeds 11 to 15 in 6 h intervals from t=0 h to t=42 h. The blue squares in panels k), m) and o) indicate the ROI used in figure \ref{fig_df_front}.}
  }
  \end{figure}

Figure \ref{fig_trans_df} panels a)-h) and i)-p) display a subset of transmission and dark-field radiographs, respectively, of germinating barley seeds 11 to 15. Radiographs in intervals of 6 h were selected starting at t=0 h and ending at t=42 h. On the grey-scale images, dark shades correspond to a high degree of respectively attenuation or scattering in the attenuation and dark-field radiographs while bright shades correspond to a low degree. 

In the transmission images in panels a)-h), the barley seeds appear homogeneous with no apparent internal changes over time. In the dark-field images in panels i)-p), however, changes over time of the internal seed structure can be observed. First, vertical dark lines are present within all of the seeds at t=0 h. Over time these lines are seen to fade until at t=30 h, all have disappeared. 

Furthermore, dark areas in the central part of the barley seeds are seen in the dark-field radiographs. In some seeds such as 11, 12 and 13 they are present at t=0 h, and in others such as 14 and 15 they seem to become more pronounced until t=18 h. As time passes, these areas shrink in all seeds. This reduction in dark-field signal begins near the scutellum part of the barley seed and is seen to spread through the endosperm region during the 42 h period of figure \ref{fig_trans_df}.

\begin{figure}[tb]
  \centering
  \includegraphics[width=0.9\textwidth]{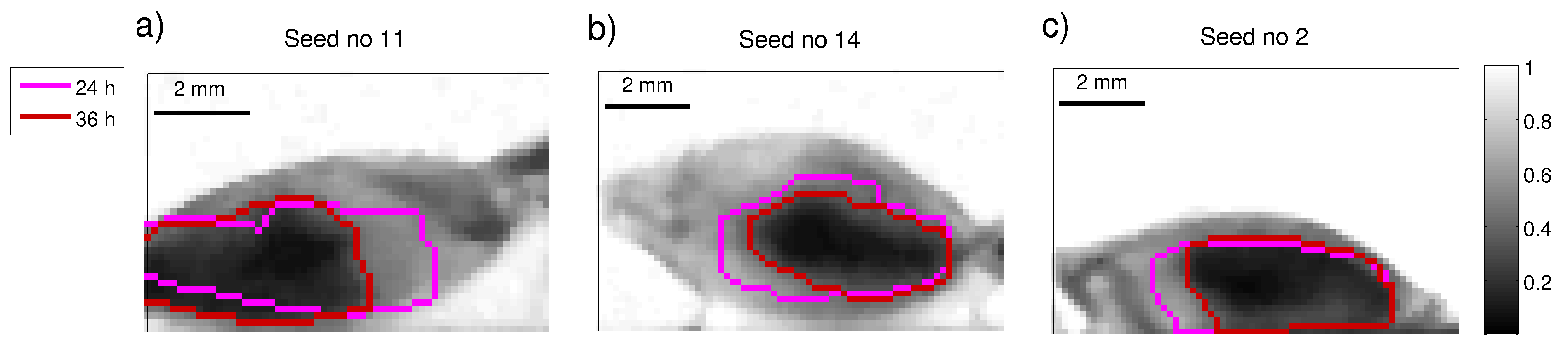}
  \caption{ \label{fig_shape}
  \footnotesize{ Illustration of the shape of the scattering signal. The perimeter at time 24 h (magenta) and 36 h (red) is shown on top of the dark-field radiograph at time 36 h. a) Seed 11. b) Seed 14. c) Seed 2.}
  }
  \end{figure}

In figure \ref{fig_shape}, three patterns of the areal change of dark-field signal in the seeds  are shown. The perimeter of the area after 36 h is highlighted in red and the perimeter at 24 h is overlayed in magenta. Panel a) shows the dominating pattern where the front of the dark-field area advances through the endosperm in parallel to the scutellar face.  This type of pattern is displayed by 16 of the 20 seeds and resembles the modification pattern from previous microscopy studies \citep{briggs_patterns_1983, briggs_malt_2002}. In panel b), a different type of pattern is seen. The dark area of high scattering diminishes from the top and bottom instead of a front in parallel to the scutellar face. This pattern was observed in seeds 4, 13, 14 and 18. In panel c), one of the two seeds that did not germinate is shown. While failing to germinate, a similar pattern to panel a) is still observed. This will be discussed further in the discussion section.

For the 16 seeds that displayed a parallel front, a quantitative analysis could be performed. To illustrate this, four snapshots of the dark-field signal of seed 11 at 12 h, 24 h, 36 h and 48 h can be seen in figure \ref{fig_df_front} panels a)-d) (for full time series see gif file in the supplementary materials). The line along the seed major axis (shown in red) and the position of the signal front (shown in magenta) were found through image analysis. In panel e), a plot of the relative position of the front along the major axis with time is shown. In the first 15 hours the front position is seen to vary around the 6.5 mm value. These variations are artefacts from the detection method, which can be sensitive towards small variations in the dark-field signal as well as changes in the seed volume due to water uptake. From 15 h to 55 h, the front position is seen to move monotonically across the seed away from the scutellar face. Similar patterns are found in the other seeds where a period of no front movement is followed by a monotonically movement away from the scutellar face. To describe this pattern, two parameters are introduced. Firstly, the initiation time will define the point in time where the front begins to move monotonically. Secondly, the speed of the front is introduced as the slope of the front position curve.

  \begin{figure}[tb]
  \centering
  \includegraphics[width=0.9\textwidth]{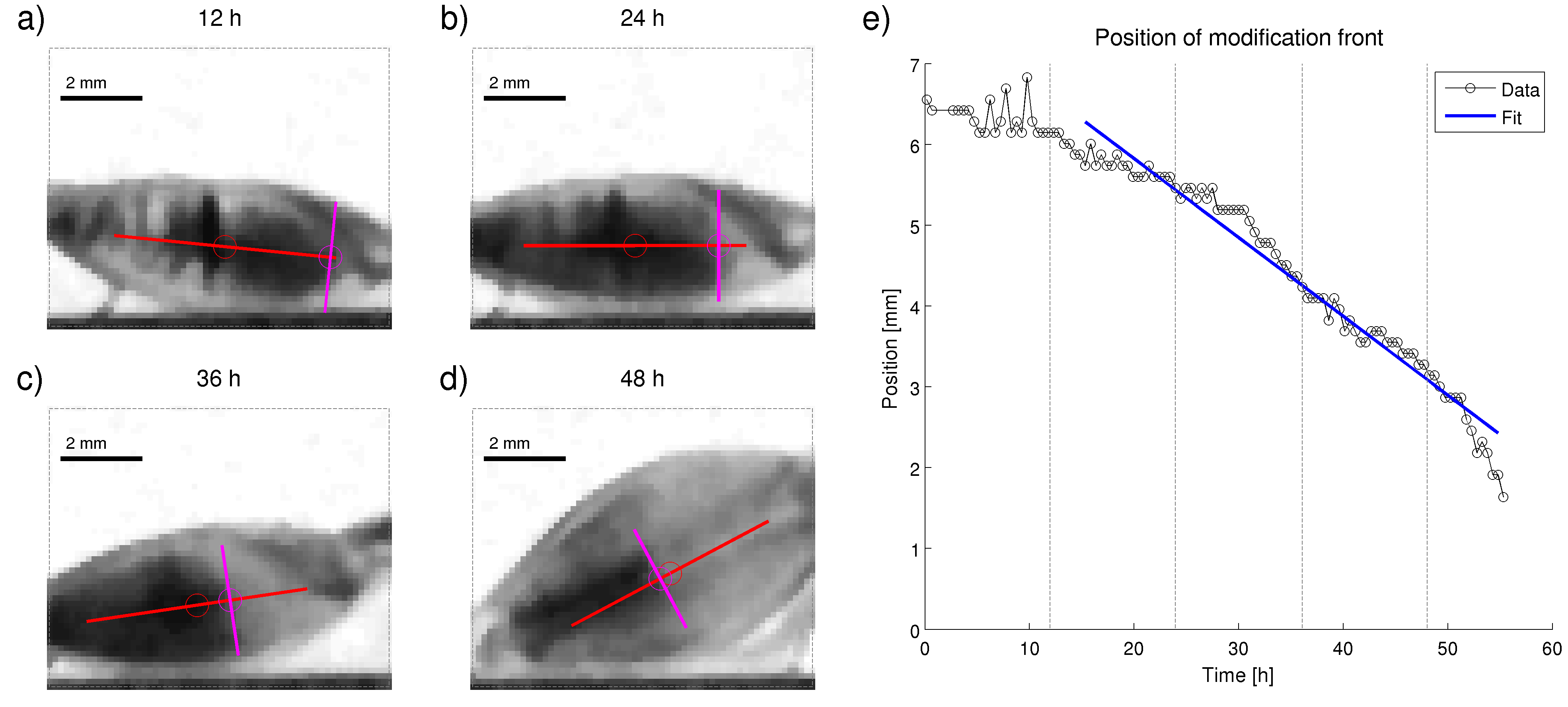}
  \caption{ \label{fig_df_front}
  \footnotesize{Image analysis of barley seeds. a)-d) Dark-field radiographs of barley seed 11 at t=12, 24, 36, 48 h. Through image analysis, the major axis and dark-field front were identified as indicated in red and magenta, respectively. e) The position of the dark-field front with time. A linear fit is indicated in blue.}
  }
  \end{figure}

To determine the initiation time a visually guided procedure is used. From the recorded front position versus time curve, the start of the monotonical movement is determined. Afterwards a visual inspection of the front in the dark-field radiographs is performed for confirmation. To find the slope of the front position curve, a first-order linear least-squares regression was used. This simple description was justified following an investigation of applying first to fourth order polynomial fits. As the investigation showed no preference towards a single polynomial order, the first-order polynomial was chosen as the simplest description. In figure \ref{fig_df_front} panel e), the resulting regression for seed 11 is shown in blue. 

\begin{table}[h!tb]
\centering
\renewcommand{\tabcolsep}{2.5mm}
\renewcommand{\arraystretch}{1.2}
\begin{tabular}{l|c|c|c|c}
Seed no&  Initiation time& Front Speed& R$^2$ (fit)& Reg. time range\\
       & [h]             & [mm/h]     &            &[h]\\
\hline
\hline     
1  & 30 & 0.13 &0.97 & 30-39\\
2 & 17 & 0.06 & 0.97 & 17-43\\
3 & 30 & 0.13 & 0.97 & 30-43\\
5 & 7  & 0.11 & 0.96 & 7-24\\
6 & 11 & 0.07 & 0.98 & 11-43\\
7 & 9  & 0.11 & 0.95 & 9-33\\
8 & 24 & 0.11 & 0.97 & 24-40\\
9 & 13 & 0.08 & 0.94 & 13-33\\
10& 28 & 0.05 & 0.91 & 28-40\\
11& 15 & 0.10 & 0.97 & 15-55\\
12& 18 & 0.08 & 0.92 & 18-55\\
15& 27 & 0.07 & 0.95 & 27-50\\
16& 33 & 0.06 & 0.88 & 33-50\\
17& 22 & 0.13 & 0.99 & 22-55\\
19&  0 & 0.07 & 0.94 & 0-22\\
20& 15 & 0.05 & 0.87 & 15-38\\
\hline
Mean & 19+/-9 &0.09+/-0.03& & \\
\hline
\end{tabular}
\caption{\label{tab_speed} Barley seed front speeds and initiation times as found through image analysis. R$^2$ values for the linear least squares regressions are indicated. No results were obtained for seed 4, 13, 14 and 18. Mean values with an interval of a single standard deviation are listed as well.}
\end{table}

Following this procedure, the initiation times and front speeds were obtained for 16 of the 20 seeds as shown in table \ref{tab_speed}. The initiation times vary from 0 h to 33 h and the speeds from 0.05 mm/h to 0.13 mm/h. Here 0 h means within the first 30 min. Also included are the R$^2$ values of the least-squares fits and the time ranges used for the regression. 

  \begin{figure}[tb]
  \centering
  \includegraphics[width=0.9\textwidth]{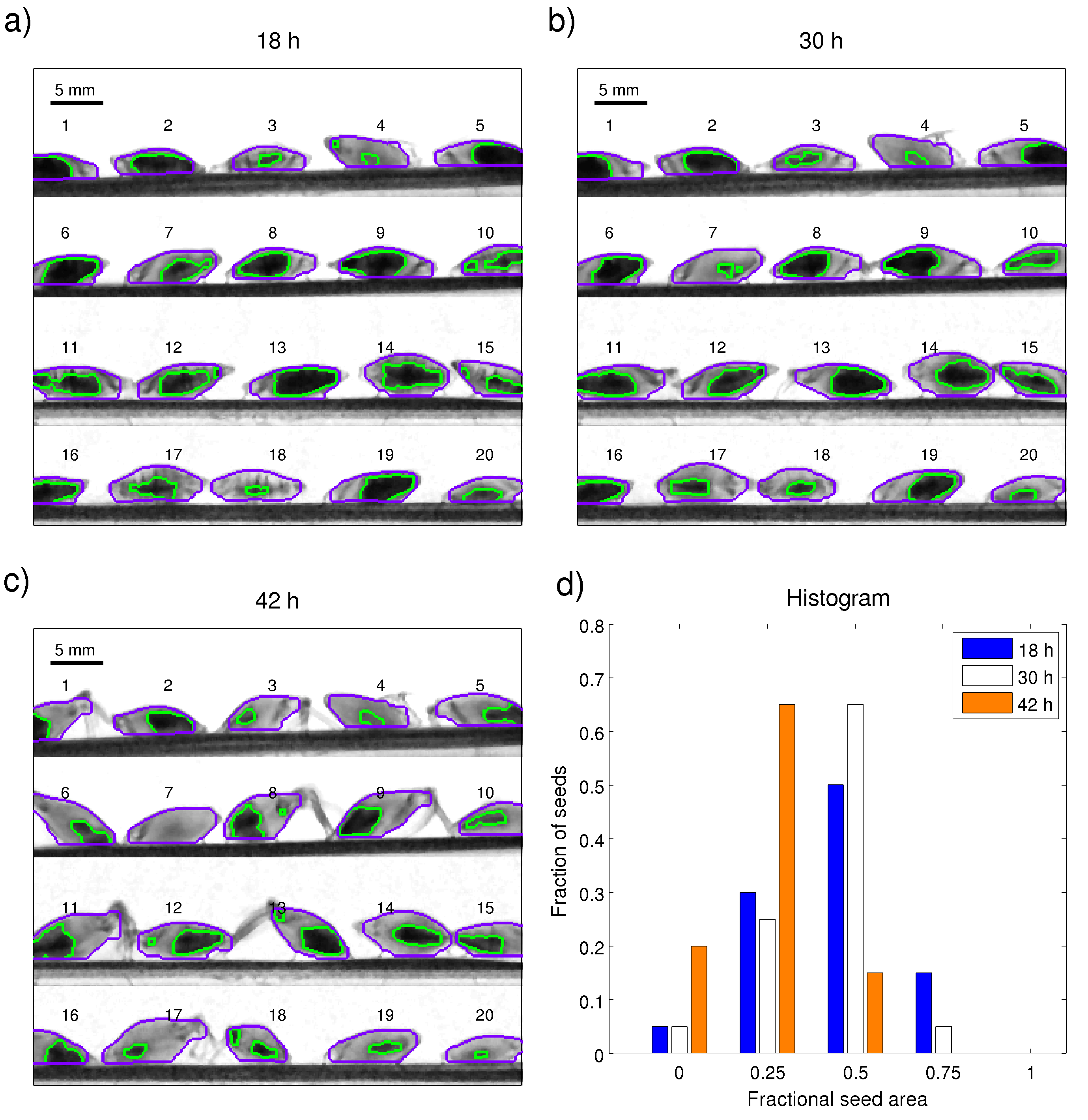}
  \caption{ \label{fig_area}
  \footnotesize{ a)-c) Dark-field radiographs for barley seeds 1 to 20 at t= 18, 30 and 42 h. From image analysis, the area of the seeds and the regions of large scattering were inditified. The perimeters of the seed and scattering areas are indicated in purple and green, respectively. d) Histogram of the ratio of the scattering areas to the full seed areas.}
  }
  \end{figure}

In figure \ref{fig_area}, the fraction of the seeds covered by dark-field signal is highlighted. On the 2D images, this will appear as an area. However, since the radiographs are acquired from the transmitted beam, the full volume contributes to the signal. In panels a)-c), dark-field radiographs of seed 1 to 20 are shown for t=12 h, 30 h and 42 h, respectively. The perimeter of the seed found by image analysis is depicted in purple, and the perimeter of the dark-field signal is shown in green. From the radiographs, it is again clear that the signal shrinks from t=18 h to t=42 h. The signal is quantified as the ratio of the dark-field area to the seed area on the radiograph. A histogram for the fractional areas are shown in panel d).

\section{Discussion}

Vertical lines as seen in the dark-field radiographs have previously been reported in studies of barley seeds using X-ray transmission radiographs \citep{demyanchuk_x-ray}.  The lines originated from stress cracks in the endosperm.  Such air-filled cracks smaller than the effective pixel size could result in a strong dark-field signal due to the electron density difference between air and seed material \citep{yashiro_unresolvable_2015}. Subsequently, filling of the stress cracks during water uptake would lead to a smaller electron density difference, and the dark-field signal would diminish as was observed. A recent study of water uptake in porous media found a similar decrease in dark-field signal when the sub-resolution pores were filled with water \citep{yang_water_porous_2014}. According to \citep{fornal_2000} stress cracks within the barley endosperm do not seem to affect the ability to germinate . 

The dark areas in the X-ray dark-field radiographs originate from X-ray scattering from microstructures in the barley seeds. With an effective $d_\textrm{GI}$ of $4 \umum$, this means that the measurements were sensitive towards microspheres with diameters in the range of $2-60 \umum$. As the starch granules are approximately spherical and fall within this range, we infer that the observed dark-field signal originated from the starch granules. Thus, the degradation of starch granules during modification would cause a reduction in signal as observed. Hence, we propose that the X-ray dark-field signal can be used to monitor the modification process. However, as the dark-field  radiographs project the scattered signal through the full 3D seed volume, a direct comparison to 2D sections from micrographs cannot be made.

Following the above argumentation, the quantitative parameters for front speed and initiation times can be related to the modification process. Assuming a linear relationship between front position and time, the front speed from the linear fit can be interpreted as a constant velocity in the spread of starch degradation as predicted from theoretical models \citep{obrien_modification_2005, fowkes_application_2010}. The recorded parameters for the seeds display a large relative variation. An uneven uptake of water from the wetted filter paper could have caused at least part of this variation. Furthermore, size, morphology and composition of the barley seed will have an influence on the water uptake and therefore the modification pattern \citep{Cozzolino_2014, molina_cano_1995}. This could explain the variations in the type of pattern as well.

Since the total dose received by the barley seeds was several Gy, radiation damage could be an issue. Indeed, this may have influenced that seed 2 and 19 failed to germinate. Although the dark-field radiographs indicated a modification process taking place, the received dose could have prevented the germination. In particular, both seeds had front speeds in the low end as seen from table \ref{tab_speed}. However, as 90\% of the seeds in the X-ray radiography measurements did germinate, radiation damage was not a critical factor overall.

\section{Conclusion}

This pilot study demonstrates the first quantitative in-situ monitoring of germinating barley seeds. Using X-ray dark-field radiography, microstructural changes within the barley seeds during water uptake and modification could be successfully monitored and assessed. 

Initially, sub-resolution stress cracks were detected in the dark-field radiographs through X-ray scattering from the air-filled pores. Water uptake in the cracks could be followed through a reduction in dark-field signal. The starch granules led to a X-ray scattering signal in the dark-field radiographs which was reduced during modification due to starch degradation. By following the changes in the dark-field signal, the modification process could be quantified by the front speed, initiation time and fractional area of un-degraded starch. In accordance with theoretical predictions, a time range existed in which the front movement could be described by a constant front speed..

Since this study represent the first non-destructive monitoring of germination processes in barley seeds, comparison of the results to indirect data from exsisting literature has proven difficult as also noted by \citep{fowkes_application_2010}. In destructive techniques such as measuring the beta-glucan content, the experimental settings are too different. Thus, further studies are needed to validate and develop the proposed method. More specifically, a study should be devised where the dark-field radiographs are compared directly to conventional methods such as micrographs.

The dose imparted on the barley seeds could be reduced in several ways. The measurement frequency could be lowered, a detector with a higher detection efficiency could reduce exposure time, and the use of a higher X-ray energy could reduce the absorption in the barley seeds. In this way, the dose per radiograph could be lowered by at least a factor of 10 and for a full time series by a factor of 50.

As the X-ray dark-field technique is not specific for barley, it could possibly also be used for investigating and monitoring the germination pattern in other types of seeds.

\section*{Acknowledgements}
      
The authors wish to thank Birthe M\o ller Jespersen for helpful discussions relating to barley. The authors are grateful to Lars Studsgaard from the Danish Malting Group for supplying the barley samples. Lastly, the authors are thankful to Keld Theodor for technical support with the experimental grating interferometer setup. The authors have received financial support from The Danish Council for Strategic Research through the NEXIM project as well as financial support from the Carlsberg Foundation.



\bibliography{xray_gbi_papers.bib,food_science_papers.bib}

      \end{document}